# A Step Test to Evaluate the Susceptibility to Severe High-Altitude Illness in Field Conditions

Eric Hermand,[1] Léo Lesaint,[2] Laura Denis,[2] Jean-Paul Richalet,[2,3] and François J. Lhuissier[2,4]

[1]Université Littoral Côte d'Opale, Université Artois, Université Lille, CHU Lille, ULR 7369—URePSSS—Unité de Recherche Pluridisciplinaire Sport Santé Société, Dunkerque, France.
[2]Université Sorbonne Paris Nord, UMR INSERM 1272 Hypoxie et poumon, Bobigny Cedex, France.
[3]Institut National du Sport de l'Expertise et de la Performance (INSEP), Paris, France.
[4]Assistance Publique-Hôpitaux de Paris, Hôpital Jean Verdier, Médecine de l'exercice et du sport, Bondy, France.

A laboratory-based hypoxic exercise test, performed on a cycle ergometer, can be used to predict susceptibility to severe high-altitude illness (SHAI) through the calculation of a clinicophysiological SHAI score. Our objective was to design a field-condition test and compare its derived SHAI score and various physiological parameters, such as peripheral oxygen saturation (SpO2), and cardiac and ventilatory responses to hypoxia during exercise (HCRe and HVRe, respectively), to the laboratory test. A group of 43 healthy subjects (15 females and 28 males), with no prior experience at high altitude, performed a hypoxic cycle ergometer test (simulated altitude of 4,800 m) and step tests (20 cm high step) at 3,000, 4,000, and 4,800 m simulated altitudes. According to tested altitudes, differences were observed in O2 desaturation, heart rate, and minute ventilation ($p < 0.001$), whereas the computed HCRe and HVRe were not different ($p = 0.075$ and $p = 0.203$, respectively). From the linear relationships between the step test and SHAI scores, we defined a risk zone, allowing us to evaluate the risk of developing SHAI and take adequate preventive measures in field conditions, from the calculated step test score for the given altitude. The predictive value of this new field test remains to be validated in real high-altitude conditions.

## Introduction

Acute mountain sickness (AMS) as well as high-altitude pulmonary edema (HAPE) and high-altitude cerebral edema (HACE) are diagnosed in field conditions essentially from clinical outcomes. However, all symptoms are nonspecific and based on subjective self-evaluation (Bärtsch and Swenson, 2013). Therefore, attempts were made to correlate the clinical outcome of these altitude-related diseases with physiological variables. The only physiological variable that can be easily assessed in field high-altitude conditions and which would likely be related to the development of AMS is pulse arterial saturation (SpO2), measurable by peripheral (fingertip) oximetry. Resting SpO2 has been associated with AMS and its severity (Karinen et al., 2010), although its measurement at rest may be influenced by some bias, such as hyperventilation due to anxiety or local cold at the fingertip. In fact, some studies failed to find a strong correlation between resting SpO2 and the occurrence of AMS (Chen et al., 2012; Wagner et al., 2012). Similarly, the risk of developing HAPE is linked to a pronounced decrease of arterial SpO2, especially when concomitant with exercise (Garófoli et al., 2010; Major et al., 2012). Tachycardia has also been associated with HAPE and might be easily measurable but is a non-specific physiological sign that can also be found in other situations such as fatigue (Bärtsch et al., 2003).

The aim of this study was then to develop an exercise step test that can be easily administered in field conditions, soon after arriving at high altitude, to predict the susceptibility to develop severe AMS or HAPE/HACE, regrouped in a clinical entity known as severe high-altitude illness (SHAI) (Canouï-Poitrine et al., 2014). Briefly, the SHAI score for an individual is calculated from various parameters such as: previous SHAI event, prior exposure to altitude, ascent speed, migraine history, habitual level of physical activity, sex, and cardiac/ventilatory response to hypoxia during exercise (Canouï-Poitrine et al., 2014).

In laboratory conditions, the hypoxic cycle ergometer test, performed on an ergocycle, has been validated to predict the susceptibility to severe forms of high-altitude-related diseases (Canouï-Poitrine et al., 2014; Richalet et al., 2021; Richalet et al., 2012). The basic principle of this test is to associate the stress of hypoxia with that of exercise to quantify the sensitivity of the chemoreceptors and therefore the ventilatory and cardiac responses to hypoxia.

Our objective was therefore to adapt the existing laboratory hypoxic cycle ergometer test to field conditions at a given altitude, using a standard step test associated with the measurement of SpO2 and HR by portable devices. Hence, the present study aims to assess the validity of a field step test compared to a laboratory hypoxic exercise test.

**Material and Methods**
**Subjects**
A group of 43 healthy subjects (15 females and 28 males) with no prior experience at high altitude participated in this study in the timeframe of a standard medical high-altitude consultation before a sojourn at high altitude, with a standard hypoxic cycle ergometer test performed on an ergocycle at a simulated altitude of 4,800 m (Richalet et al., 2021). Mean age was 30.6 ± 11.4 years, height was 174 ± 8.0 cm, body weight was 68.5 ± 11.7 kg, and body mass index was 22.6 ± 2.7 kg.m$^{-2}$.

The protocol was approved by the "Comité de Protection des Personnes du Sud-Ouest et Outre-Mer IV" Ethics Committee (Registration number: 2017-A00986-47). All participants provided written informed consent before participation.

**Experimental design**
After the initial standard evaluation, all participants were invited to perform, in a random order, four step tests at various simulated altitudes: 0 m (ambient air, baseline), 3,000 m, 4,000 m, and 4,800 m; each of them followed by a 30-minute rest phase.

The standard hypoxic cycle ergometer test consists of four successive phases, on a cycle ergometer, of 3–4 minutes each with the following sequence: rest in normoxia, rest in hypoxia, exercise in hypoxia (at an intensity aiming for 40%–50% of HR reserve), and exercise in normoxia at the same intensity, (Richalet et al., 2021; Richalet et al., 1988). The step test consisted of stepping up and down a 20-cm high step, one leg after another, with a frequency of 90 steps/min for 6 minutes, enforced with a metronome. On this matter, based on preliminary HR and SpO2 data observed during exercise (data not presented), a 20 cm high step test allows subjects to reach similar exercise intensities at high altitude that the hypoxic cycle ergometer test: a 40 cm high step, which is the regular step height of the Harvard step test for male subjects (Ryhming, 1953), leads to a too vigorous exercise intensity, potentially difficult to sustain during 6 minutes for some people, especially at 4,800 m. Moreover, 20 cm remains a convenient step height on the field because this is the average stairs step height.

Minute ventilation ( E, L.min−1, Vyntus CPX, Carefusion, San Diego, CA, USA), (HR, bpm, ECG Cardiosoft, GE, Boston, MA, USA), and SpO2 (pulse oximetry with prewarmed ear lobe sensor, %, Nonin Medical Inc., Plymouth, MN, USA) were continuously monitored. During the step tests, a fingertip oximeter (Wrist Ox2, Nonin Medical Inc., Plymouth, MN, USA), was added in order to simulate field conditions and allow the measurement of HR and fingertip SpO2 (SpftO2).

Altitudes were simulated by a gas mixture apparatus (Altitrainer, SMTEC, Nyon, Switzerland) allowing breathing via a face mask a O2/N2 gas mixture with various FIO2 (0 m, 20.95%; 3000 m, 14.5%; 4000 m, 12.7%; 4800 m, 11.5%).

Standard variables of ventilatory and cardiac responses to hypoxia were calculated from E, HR, and SpO2 data continuously measured and then averaged on the last minute of each phase during the hypoxic cycle ergometer test and the step tests: exercise-induced desaturation in hypoxia (ΔSpO2e = SpO2EN − SpO2EH, with SpO2EN and SpO2EH being SpO2 measured during exercise in normoxia and in hypoxia, respectively), ventilatory response to hypoxia during exercise [HVRe = ( E−EH −  E−EN)/(ΔSpO2e × body weight) × 100, with E-EH and E-EN being E measured during exercise in hypoxia and in normoxia, respectively], cardiac response to hypoxia during exercise [HCRe = (HREH − REN)/ΔSpO2e, with HREH and HREN being HR measured during exercise in hypoxia and normoxia, respectively].

Values for the step tests were obtained from the differences between the step test at a given altitude and the step test performed in normoxia. The SHAI score was calculated from physiological and clinical data, as previously described (Canouï-Poitrine et al., 2014; Lhuissier et al., 2012). The step test score was calculated with the same items as the SHAI score except HVRe (which cannot be measured during the step test in the field conditions). Items of the SHAI and step test scores for persons without experience of high altitude are given in Table 1. A linear regression was then performed to quantify the relation between step test score at each altitude and the standard SHAI score, using the following equation:

Step test score (altitude) = a(altitude) x SHAI score + b(altitude)

**Table 1**
Computation of SHAI and step test scores to define the individual susceptibility to Severe High-Altitude Illness (SHAI) in subjects without previous experience at high altitude. (adapted from (Richalet et al., 2021)).

|  | SHAI score | step test score |
|---|---|---|
| **Planned daily altitude gain (>400 m/night)** | 2 | 2 |
| **Geographical location (Aconcagua, Ladakh-Zanskar, Mont-Blanc)** | 0.5 | 0.5 |
| **Female sex** | 0.5 | 0.5 |
| **Regular endurance physical activity*** | 1 | 1 |
| **HVRe (L.min$^{-1}$.kg$^{-1}$) < 0.68** | 3 | |
| **HVRe (L.min$^{-1}$.kg$^{-1}$) ≥ 0.68 and < 0.94** | 1 | |
| **HCRe (b.min$^{-1}$.%$^{-1}$) < 0.72** | 1 | 1 |
| **HCRe (b.min$^{-1}$.%$^{-1}$) ≥ 0.72 and < 0.95** | 1 | 1 |
| **ΔSpO$_2$e (%) ≥ 24** | 2 | 2 |
| **ΔSpO$_2$e (%) ≥ 19 and < 24** | 1 | 1 |
|  |  |  |
| **Threshold to define high susceptibility** | > 5.5 | |

HVRe: ventilatory response to hypoxia at exercise
HCRe: cardiac response to hypoxia at exercise
ΔSpO$_2$e: decrease in arterial O$_2$ saturation in hypoxia at exercise
*At least 40 min intense aerobic exercise 3 times/week

In this linear equation, a and b are given for each altitude of step test and can be determined for any given altitude. Knowing that the threshold for the SHAI score is 5.5, beyond which a given individual is likely to be intolerant to high altitude, a risk zone is calculated so these subjects can determine at a given altitude if she/he is at risk considering the step test score calculated at this altitude.

**Statistical analysis**

All variables are expressed as mean ± standard deviation. Comparison of variables between the hypoxic cycle ergometer test and the step tests was performed using a one-way analysis of variance (ANOVA) with repeated measures, followed by a Bonferroni adjusted post hoc test (Bonferroni method). A two-way ANOVA was performed to compare the effects of altitude and the method of measurement of HR (ECG vs. fingertip) and SpO2 (earlobe vs. fingertip). The statistical significance was set at $p < 0.05$.

**Results**

Values of main physiological variables are presented in Table 2.

**Table 2**

Physiological variables during the standard hypoxia exercise test and the step tests performed at various altitudes. $\Delta SpO_2e$: desaturation during exercise. HRe: heart rate at exercise, $\dot{V}_{Ee}$: minute ventilation at exercise; HCRe: cardiac response to hypoxia at exercise; HVRe: ventilatory response to hypoxia at exercise. SHAI score for the hypoxic exercise test, step test score for the step tests calculated at each altitude, without taking into account HVRe.
Compared to hypoxic exercise test: *p<0.05, ***p<0.001

|  | Hypoxic exercise test | Step test 3000 m | Step test 4000 m | Step test 4800 m | P |
|---|---|---|---|---|---|
| $\Delta SpO_{2e}$ (%) | 24.8 ± 4.6 | 13.2 ± 2.9 *** | 22.5 ± 3.2 *** | 30.2 ± 3.9 *** | <0.001 |
| $HR_e$ (bpm) | 129.0 ± 12.8 | 122.4 ± 17.4 *** | 131.6 ± 17.8 | 138.7± 17.1 *** | <0.001 |
| $\dot{V}_{Ee}$ (L.min$^{-1}$) | 38.2 ± 7.7 | 35.6 ± 6.8 * | 38.0 ± 6.0 | 41.1 ± 8.0 * | <0.001 |
| $HCR_e$ (bpm.%$^{-1}$) | 0.92 ± 0.31 | 1.08 ± 0.55 | 1.01 ± 0.28 | 0.99 ± 0.24 | 0.075 (ns) |
| $HVR_e$ (L.min$^{-1}$.kg$^{-1}$) | 0.59 ± 0.31 | 0.54 ± 0.40 | 0.50 ± 0.29 | 0.50 ± 0.23 | 0.203 (ns) |
| SHAI or step test score | 4.81 ± 1.67 | 0.92 ± 0.85 *** | 2.02 ± 0.80 *** | 2.70 ± 0.75 *** | <0.001 |

**Table 3**

Heart rate (HR) and peripheral oxygen saturation (SpO$_2$) at rest (r) and exercise (e) measured with EKG + ear lobe oximeter (EKG) and fingertip oximeter (foxi).
Compared to step test at 0 m: ***, p<0.001
EKG + ear lobe vs. finger oximeter: ###, p<0.001

|  |  |  | Step test 0 m | Step test 3000 m | Step test 4000 m | Step test 4800 m | Altitude effect (P) | Method effect (P) |
|---|---|---|---|---|---|---|---|---|
| SpO$_2$ | Rest | SpO$_2$r EKG | 98.9±0.1 | 94.0±0.4 *** | 90.2±0.4 *** | 86.3±0.5 *** | <0.001 | <0.001 |
|  |  | SpO$_2$r foxi | 97.3±0.2 ### | 91.0±0.4 ###*** | 87.0±0.5 ###*** | 83.0±0.6 ###*** |  |  |
|  | Exercise | SpO$_2$e EKG | 99.0±0.1 | 85.8±0.5 *** | 76.6±0.6 *** | 68.5±0.6 *** | <0.001 | <0.001 |
|  |  | SpO$_2$e foxi | 96.6±0.2 ### | 82.2±0.4 ###*** | 74.1±0.5 ###*** | 68.6±0.6 *** |  |  |
| **HR** | Rest | HRe EKG | 108.9±2.6 | 122.4±2.7 *** | 132.1±2.7 *** | 140.0±2.7 *** | <0.001 | ns |
| | | HRe foxi | 112.8±3.1 | 125.3±3.3 *** | 127.6±2.9 *** | 136.2±3.5 *** | | |
| | Exercise | HRr EKG | 72.4±1.7 | 78.8±1.8 *** | 80.7±1.9 *** | 83.3±1.9 *** | <0.001 | ns |
| | | HRr foxi | 72.2±1.7 | 79.0±1.9 *** | 79.8±1.9 *** | 82.7±2.0 *** | | |

As expected, desaturation during exercise (decrease in SpO2), as well as HR and E, increases with altitude. The values measured during the step test at 4,000 m were not different from values measured at 4,800 m during the hypoxic cycle ergometer test. Values of HCRe and HVRe were similar in all conditions.

Values of HR and SpO2 at rest (r) and exercise (e) measured with ECG and ear lobe oximeter during the standard hypoxic exercise test and measured with the fingertip oximeter are given in Table 3.

SpO2 measured with fingertip oximeter was always lower than SpO2 measured at the prewarmed ear lobe (Table 3, p < 0.001).

Values of step test score at various altitudes as a function of SHAI score calculated during the hypoxic cycle ergometer test are shown in Figure 1. Linear relationships were computed between the step test scores and the hypoxic cycle ergometer test score.

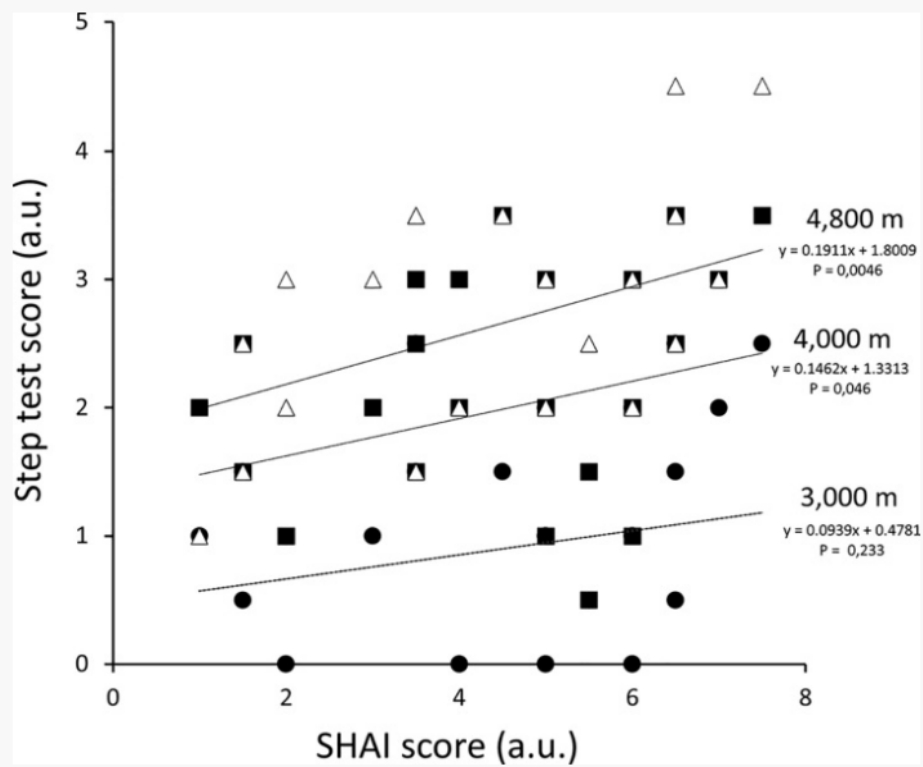

FIG. 1. Values of step test score as a function of severe high-altitude illness (SHAI) score at 3,000 m (black dots), 4,000 m (black squares) and 4,800 m (white triangles). Linear regressions are calculated for each altitude. For 3,000 m: r = 0.186 (ns); 4,000 m: r = 0.305 (p < 0.05); 4,800 m: r = 0.420 (p < 0.01).

Values of a and b calculated for each altitude are given in Figure 2, with the linear regression equations.

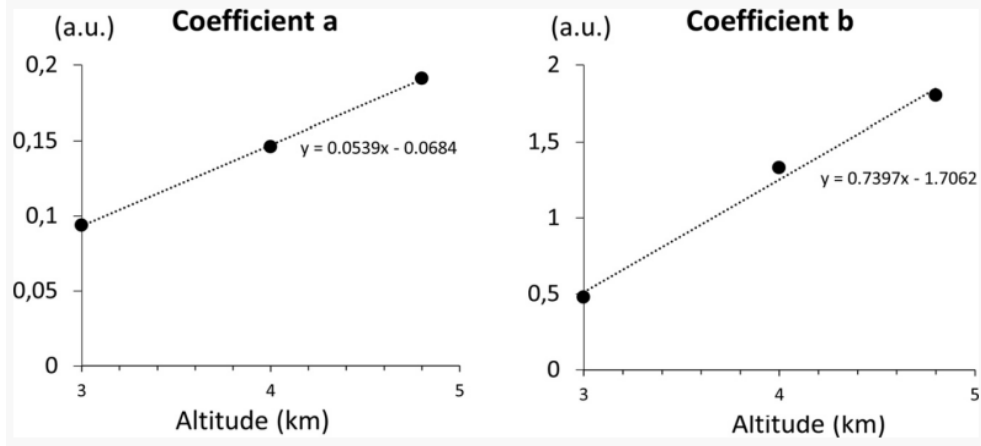

FIG. 2. Left panel (a): values of a (coefficient of Eq. 1) for 3, 4, and 4.8 km altitudes. Right panel (b): values of b (ordinate at the origin of equation 1) for 3, 4, and 4.8 km of altitude. Linear regression lines are shown. For (a): r = 0.99 (p < 0.001); (b): r = 0.99 (p < 0.001).

From values of a and b at a given altitude, we calculated the expected step test score for a given SHAI score and a given altitude (in km):

$$\text{step test score} = \text{SHAI score} \times (0.0539 \times \text{altitude} - 0.0684) + 0.7397 \times \text{altitude} - 1.7062 \quad (\text{Equ.2})$$

Considering 5.5 as the threshold for the SHAI score above which the subject is at risk (Canoüi-Poitrine et al., 2014; Richalet et al., 2021), we calculated the limit line from Equ. 2:

$$\text{step test score} = 5.5 \times (0.0539 \times \text{altitude} - 0.0684) + 0.7397 \times \text{altitude} - 1.7062 \quad (\text{Equ. 3})$$

$$\text{step test score} = 1.03615 \times \text{altitude} - 2.0824 \quad (\text{Equ. 4})$$

We then defined a "risk zone" and a "safe zone" above and under this limit line, respectively (Fig. 3).

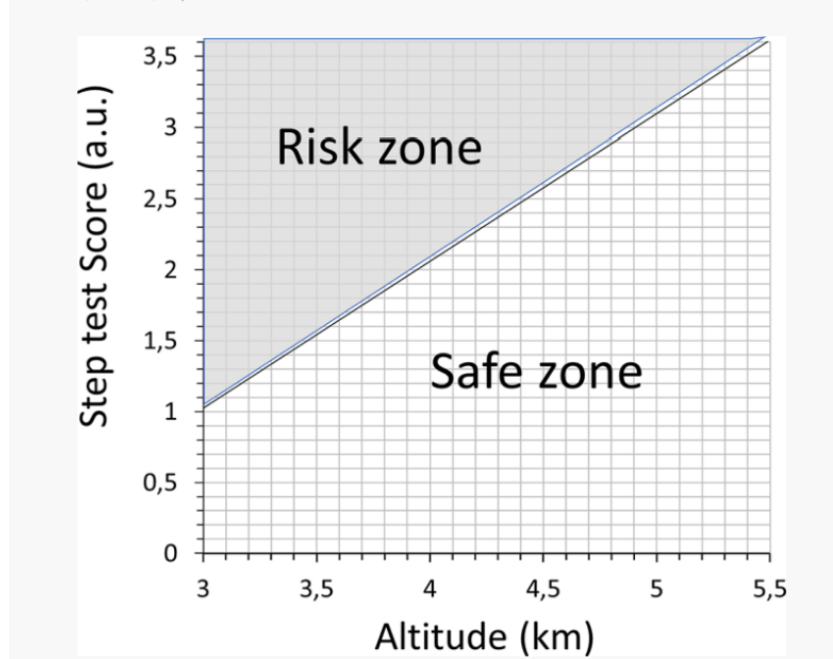

FIG. 3. Limit line computed for a SHAI score of 5.5 for a given altitude and the corresponding step test score. Above this line, the subject is considered at risk and should take preventive measures (rest, acetazolamide treatment, descent, …).

**Discussion**

Our aim was to validate an easy-to-use field test allowing people going to high altitude to evaluate their personal risk to develop SHAI. We described a test that needs subjects to perform two 6-min step tests, the first one at sea level before going to altitude, as a baseline test, and the second one as soon as they arrive at an altitude from 3,000 to 4,800 m, before the acclimatization process. During the two step tests, subjects need to measure their HR and SpO2 with a pulse oximeter in order to calculate a step test score including clinical and physiological ($\Delta SpO2e$ and HCRe) data. Identifying his step test score at a given altitude will allow a subject to classify himself/herself in a "safe zone" or in a "risk zone" where he/she should take SHAI prevention measures.

The standardized laboratory hypoxic cycle ergometer test used to determine a SHAI score is performed on an ergocycle with an exercise intensity of 40%–50% of HR reserve and a 4,800 m simulated altitude.

Cardiac and ventilatory responses to hypoxia, i.e., HR and E differences between normoxia and hypoxia related to desaturation during exercise, are not modified when subjects perform ergocycle or step tests, whatever the simulated altitude. We previously described that those two variables were not modified when the test was performed on an ergocycle at different altitudes and different exercise intensities (Lhuissier et al., 2012). These new data demonstrate that these variables are robust and independent of altitude, exercise modality and exercise intensity. They are inherent to the individual chemosensitivity to hypoxia.

The calculation of the step test score does not take into account HVRe, an important item of the SHAI score to estimate the individual susceptibility to SHAI. This could be interpreted as a limitation of the step test score but minute ventilation is not easy to measure in field conditions. An easy-to-use score has to be based on clinical and physiological data that can be easily measured. Till date, HR and SpO2 are two physiological variables easily measurable on the field and are part of the step test score.

Cobb et al. used a step-test at altitude called "Xtreme Everest Step-Test" to predict AMS on a trek to Everest base camp (Cobb et al., 2021). This step test was performed with the same step height of 20 cm but lasted only 2 minutes at a 60/min step rate. They showed that a lower SpO2 during the step test performed at 3,500 m predicted the development of moderate-to-severe AMS during the ascent. This is in accordance with our results, as a greater $\Delta SpO2e$ (and thus a lower SpO2 during the step test at altitude) leads to a higher step test score.

Concerning the step test modalities, we performed preliminary step tests, following the work by Rhyming (1953): step height 40/33 cm for men/women, step cycle cadence 22.5/min, duration 5 minutes; which was itself adapted from the princeps article by Brouha (1943): step height 50 cm, cadence 30/min, duration 5 minutes). Regarding these data and our field-related constraint, in an ecological approach, we finally settled for a lower step height (20 cm), commonly met in public buildings or hotels/houses. As a result, compared with data collected during the laboratory hypoxic cycle ergometer test at 4.800 m, we observed a greater desaturation (30.2% ± 3.9% vs. 24.8% ± 4.6%, Table 1) and a greater HR at exercise (138.7 ± 17.1 vs. 129.0 ± 12.8 bpm) during the step test conducted at 4,800 m, but similar desaturation (22.5% ± 3.2% vs. 24.8% ± 4.6%) ant HR (131.6 ± 17.8 vs. 129.0 ± 12.8 bpm) during the step test conducted at 4.000 m. This is particularly interesting, as several high-altitude treks start at an altitude between 3,000 m and 4,000 m, qualifying the step test as a valid protocol to assess the susceptibility to SHAI.

**Limitations**

The subjects in this study were overall "low responders" to hypoxia since their HVRe was <0.68 l.min$^{-1}$·kg$^{-1}$ in 30 subjects and the SHAI score was >5.5 in 25 of 43 subjects. Our observations should then be confirmed in "high responders" subjects. In the same way, our subjects were all unexperienced with high altitude. As our conclusions may not be extended to regular high-altitude trekkers yet, this simple field test would still provide some useful first-hand information to inexperienced travelers about potential harmful effects of altitude.

Moreover, the test might not be useful or valid if acetazolamide was taken as a prophylactic drug before arrival to altitude.

Another limitation is the validity of a wearable pulse oximeter. We showed than SpO2 measured with a fingertip oximeter is systematically lower than SpO2 measured at the prewarmed ear lobe. A correcting factor could then be applied in the absence of a reliable oximeter. In the same manner, the integration of SpO2 measurement in recent smartwatches must be used with caution, as their reliability and accuracy are still to be demonstrated (Hermand et al., 2021).

**Perspectives**

The step test score should be validated in subjects going to real altitude and by performing the test in field conditions. An assessment of AMS symptoms, for example, with the Lake Louise score (Hackett and Oelz, 1992), in these subjects should be made in order to directly correlate the step test score with the occurrence of AMS.

**Conclusion**

The computation of the step test score at any altitude between 3,000 and 4,800 m could help evaluate the individual risk of developing SHAI and allow taking preventive measures such as limiting exercise, rest, acetazolamide use, and limiting the speed of ascent or even descent to lower altitudes. It requires a simple equipment—a fingertip oximeter and a flat-surfaced 20 cm high object (stairs, rock, box)—to perform a 6-min step test. It should be validated in a field study.

**Authors' Contributions**

E.H., L.L., L.D., J.P.R., and F.L. were involved in data collection, data analysis, and article drafting. E.H., J.P.R., and F.L. were involved in study design and overview. All authors were involved in final revision and approval of the article.

**Author Disclosure Statement**

None of the authors have any relevant conflicts of interest to disclose.

**Funding Information**

No funding was received for this article.